\documentclass[10pt]{article}
\usepackage[dvips]{graphicx}
\usepackage{amssymb}
\begin{document}
%\twocolumn{}
\begin{center}
{ \LARGE {\bf Construction of a few Quantum mechanical Hamiltonians via L$\acute{e}$vi-Leblond type Linearization:Spinor states and Supersymmetry.\\\
 } }
\end{center}
\vskip 0pt
\begin{center}
{\it { $Arindam\hskip 2 pt Chakraborty\footnote{arindam.chakraborty@heritageit.edu}$\\
Bhaskar\hskip 2 pt Debnath\\
Ritaban\hskip 2 pt Datta\\
\it {Department of Physics,Heritage Institute of Technology,Kolkata-107,India}  }}\\
{\it{$Pratyay\hskip 2 pt Banerjee$\\
P. R. Thakur Government College, Thakur Nagar, 743287, West Bengal, India }}
\end{center}

\vskip 20pt
\begin{center}
{\bf Abstract}
\end{center}

\par A number of new L$\acute{e}$vi-Leblond type equations admitting four component spinor solutions have been proposed. The pair of linearized equations thus obtained in each case lead to Hamiltonians with characteristic features like L-S coupling and supersymmetry. The relevant momentum operators have often been understood in terms of Clifford algebraic bases producing Schr$\ddot{o}$dinger Hamiltonians with L-S coupling. As for example Hamiltonians representing Rashba effect or three dimensional harmonic oscillator have been constructed. The supersymmetric nature of one dimensional oscillator has also been appreciated.

 \par PACS Number(s):  05.45.Pq, 05.45.Ac, 05.45.-a.

\section{Introduction}
The appearance of spin as quantum observable and spinor states associated with it have very often been attributed to Dirac theory of relativistic electron. In a seminal paper $^{1}$ Dirac considered the Lorentz invariant Klein-Gordon equation as a representative equation of a free electron in a special relativistic frame work. The need for a consistency in the physical interpretation of negative energy and probability density along with the criterion of Lorentz invariance led Dirac to propose a kind of linearized equation i.e. ; 1st order in space and time $^{2, 3}$. The said linearization involves matrix coefficients (the celebrated gamma matrices) and hence leads to states represented by column vectors known as spinors. One of the prime consequences of this kind of construction motivated by special theory of relativity is the emergence of spin as quantum mechanical observable and a necessary object to be fed up with orbital angular momentum in order to have a conservation principle of total angular momentum.
The very idea of spin as an outcome of a Lorentz invariant theory (special theory of relativity) has been questioned by Mark J. L$\acute{e}$vy-Leblond$^{4}$ who considered free particle Schrodinger wave equation in the light of Galilean invariance. An attempt to linearize the Schrodinger equation leads again to a pair of equations involving four component spinors. The method can be reversed i. e. ; one can start from a couple of linearized equations involving spinor solutions and end up with a Schrodinger equation admitting spinor states$^5$. The success of this method is two fold:
(i)	It can be extended from free particle Hamiltonian to the Hamiltonian for a particle in electromagnetic field involving a generalized momentum taking magnetic vector potential into regard. The Pauli term that has otherwise been introduced heuristically in the Schrodinger equation comes naturally in this formulation.
(ii)	It holds the very process of linearization responsible for the existence of spin in the relevant theory, not the spin as a consequence of purely relativistic effect,  a fact that is commonly believed and frequently escapes critical attention.
The method has been appreciated by C. R. Hagen$^{6}$ to construct three dimensional harmonic oscillator with L-S coupling and multi-component spinor solution has been proposed.
\par In the present report we consider some possible extensions of the method proposed by L$\acute{e}$vy-Leblond. We have taken up a number of examples of Hamiltonian of both theoretical and practical interest and shown how they can be constructed from a pair of linearized equations involving four component spinors, each involving both the two component states. For a problem with so called scalar potential it is not always possible to reduce each of the two component states obeying their corresponding Schr$\ddot{o}$dinger equation because one member of the pair remains merely as a constraint relation$^{6}$ for the system. But for some potential the free particle momenta have been replaced by some non hermitian gauge momentum like objects and their complex conjugate. The said Hamiltonians can be constructed by representing the generalized momentum operators in Clifford bases$^{7}$. The physical inferences thus obtained are highly nontrivial. For example in one such case we obtain an extended version of Rashba Hamiltonian $^{8, 9}$from which the Rashba type spin-momentum term appears as special case. We also obtain the Dresselhaus Hamiltonian by choosing gauge momenta that involve Pauli matrices. Rashba and Dresselhaus model have been studied widely to understand  the behaviour of various physical systems ranging from semiconductor heterostructures $^{10}$ to superconductivity $^{11}$. A few observations are interesting in this regard: (i)	With this method it is possible to construct a number of one particle Hamiltonians in three dimension with constant or spatially varying spin orbit coupling terms. The choice of Pauli matrices as Clifford bases comes out to be responsible for spin orbit coupling terms appearing in the Hamiltonians in majority of the occasions.(ii)	In some cases we observe the presence of  certain interaction terms with opposite sign in the Hamiltonians acting on two different two component spinors. For example the case with one dimensional harmonic oscillator our search for spinor like states eventually leads to the supersymmetric quantum mechanics with a pair of Hamiltonians corresponding to bosonic and fermionic sectors$^{12}$.(iii)	While discussing L$\acute{e}$vi-Leblond scheme we also comment that the process of linearization is not unique at least to within a loop algebraic context involving Pauli matrices. The physical meaning of all such extensions may not be so much obvious at present.(iv)	Finally, the existence of generalized momentum corresponding to various interactions and space dependent potentials is reminiscent of Hertz’s contention in classical mechanics regarding the kinetic origin of potential energy$^{13, 14}$. The said linearization process once again elaborates this fact in a number of physical theories.

\section{L$\acute{e}$vy-Leblond scheme}

A typical Schr$\ddot{o}$dinger eigenvalue problem for a free particle can be viewed as a
\begin{eqnarray}
\frac{P^2}{2}\bf {1_{4\times 4}}\bf{\Psi}=E\bf{\Psi}
\end{eqnarray}
Where $P_j=-i\partial_j$ is the linear momentum operator and $\bf{\Psi}=(\psi_1, \psi_2, \eta_1, \eta_2)^T$ can be viewed as a 4-component vector solution of the said equation. Identifying $\psi=(\psi_1, \psi_2)^T$ and $\eta=(\eta_1, \eta_2)^T$ it is possible to obtain a pair of coupled linear equations of $\psi$ and $\eta$ via L$\acute{e}$vi-Leblond scheme. The involvement of Pauli matrices $\{\sigma_j\vert j=1, 2, 3\}$ in the linearized equation thus obtained gives $\Psi$ the status of a four component spinor while each of the components corresponds to the same energy eigenvalue $E$. In orde rto effect the scheme let us write
\begin{eqnarray}
(A^{\prime}E+B_i^{\prime}P_i+C^{\prime})(AE+B_jP_j+C)=2E-P_kP_k
\end{eqnarray}
leading to the relations
\begin{eqnarray}
A^{\prime}A=0\nonumber\\
A^{\prime}C+C^{\prime}A=2\nonumber\\
C^{\prime}C=0\nonumber\\
A^{\prime}B_j+B_j^{\prime}A=0\nonumber\\
B_i^{\prime}B_j+B_j^{\prime}B_i=-2\delta_{ij}\nonumber\\
C^{\prime}B_i+B_i^{\prime}C=0
\end{eqnarray}
Identifying $B_4=i(A+\frac{1}{2}C)$, $B_4^{\prime}=i(A^{\prime}+\frac{1}{2}C^{\prime})$, $B_5=A-\frac{1}{2}C$, $B_5^{\prime}=A^{\prime}-\frac{1}{2}C^{\prime}$

the above relation can be represented in a condensed form
\begin{eqnarray}
B_i^{\prime}B_j+B_{j}^{\prime}B_i=-2\delta_{ij}
\end{eqnarray}
By considering an invertible matrix $\Lambda$ one can write $B_5=-i\Lambda$, $B_5^{\prime}=-i\Lambda^{-1}$, $B_k=\Lambda\gamma_k$ and $B_k^{\prime}=-\gamma_k\Lambda^{-1}$ consistent with equation-4. The $\gamma$-matrices are defined as
\begin{eqnarray}
\gamma_{j}= \left(\begin{array}{cc}
   0 & \sigma_j \\
   \sigma_j & 0
    \end{array} \right)
\end{eqnarray}
and
\begin{eqnarray}
\gamma_4= \left(\begin{array}{cc}
   1 & 0 \\
   0 & -1
    \end{array} \right)
\end{eqnarray}
Choosing
\begin{eqnarray}
\Lambda= \left(\begin{array}{cc}
   0 & 1 \\
   1 & 0
    \end{array} \right)
\end{eqnarray}
Hence calculating $A, B, C$ matrices we get the following equation
\begin{eqnarray}
\sigma_jP_j\psi+2i\eta=0\nonumber\\
\sigma_jP_j\eta-iE\psi=0
\end{eqnarray}
As the $\sigma$ matrices provide the bases of Clifford algebra$(cl_3)$ the operator $P$ can be written in general in terms of Clifford algebraic bases $(e_1, e_2, e_3)$ holding the relations
\begin{eqnarray}
e_j^2=1\nonumber\\
e_ie_j+e_je_i=0
\end{eqnarray}
Hence writing $\wp=P_!e_1+P_2e_2+P_3e_3$ as Clifford $1$-vector operator one can rewrite the L$\acute{e}$vi-Leblond equation
\begin{eqnarray}
\wp\psi+2i\eta=0\nonumber\\
\wp\eta-iE\psi=0
\end{eqnarray}
It is easy to verify that $\psi$ and $\eta$ as two component vectors separately satisfy free particle Schr$\ddot{o}$dinger equation like the following
\begin{eqnarray}
\wp^2\psi=E\psi\nonumber\\
\wp^2\eta=E\eta
\end{eqnarray}
\par Our present method is to reverse the process starting from a suitable linearized equation of L$\acute{e}$vy-Leblond type and going back to Schr$\ddot{o}$dinger equation involving various Hamiltonians of physical relevance. The choice of Clifford elements motivates the possibility of such linearization using matrices other than Pauli matrices. For example the same free particle Hamiltonian can be arrived using a loop algebra isomorphic to $su(2)$ with bases $(\sigma_j\bigotimes K)$, $j=1, 2, 3$ and $K$ is any idempotent. It can be done even by non-hermitian matrices. This method has been applied$^{4, 5}$ to obtain Pauli-Schr$\ddot{o}$dinger Hamiltonian where the Pauli term appears naturally unlike the heuristic way as has been introduced by Pauli.
\section{Construction of Hamiltonians}
Considering an operator which can be called generalized Clifford momenta as  $\wp^A=e_j(P_j+A_j)$ and $\wp^B=e_j(P_j+B_j)$ in Clifford bases  a couple of linearized equations can be given as
\begin{eqnarray}
\wp^A\psi+2i\eta=0\nonumber\\
\wp^B\eta-iE\psi=0
\end{eqnarray}
With the rule of derivation $\wp\doteq\wp\rfloor +\wp\wedge$ eqn leads to Schr$\ddot{o}$dinger equations
\begin{eqnarray}
H^{AB}\psi=E\psi\nonumber\\
H^{BA}\eta=E\eta
\end{eqnarray}
or simply
\begin{eqnarray}
H\Psi=E\Psi
\end{eqnarray}
where the $4\times 4$ matrix Hamiltonian is given by
\begin{eqnarray}
H= \left(\begin{array}{cc}
   H^{AB} & 0 \\
   0 & H^{BA}
    \end{array} \right)
\end{eqnarray}
Where
\begin{eqnarray}
H^{AB}=\wp^A\wp^B\nonumber\\
H^{BA}=\wp^B\wp^A
\end{eqnarray}
Choosing Pauli matrices to be Clifford elements

\begin{eqnarray}
\frac{1}{2}[P^2+(P_jA_j+B_jP_j+B_jA_j)+i\epsilon_{ijk}\sigma_i(P_jA_k+B_jP_k+B_jA_k)]\psi=E\psi
\end{eqnarray}
\begin{eqnarray}
\frac{1}{2}[P^2+(P_jB_j+A_jP_j+A_jB_j)+i\epsilon_{ijk}\sigma_i(P_jB_k+A_jP_k+A_jB_k)]\eta=E\eta
\end{eqnarray}
\subsection{Case-1}

Choosing $A_j=-i\alpha_j=B^*_j$ and $(\alpha_1\alpha_2\alpha_3)\in R^3$ is a constant.

\begin{eqnarray}
 \left(\begin{array}{cc}
   H_0-H_{\alpha} & 0 \\
   0 & H_0+H_{\alpha}
    \end{array} \right)\Psi=E\Psi
\end{eqnarray}
where $H_{\alpha}=\epsilon_{ijk}\sigma_i\alpha_jP_k$ and $H_0=\frac{P^2+\alpha^2}{2}$. For $\alpha_3=\beta$ and $\alpha_1=\alpha_2=0$ we can recover Rashba interaction term
\begin{eqnarray}
H_{\alpha}=\beta(\sigma_1P_2-\sigma_2P_1)
\end{eqnarray}
\subsection{Case-2}
Choosing $A=\frac{\alpha}{2} (\sigma_1, -\sigma_2)=\pm B$ one can obtain Hamiltonian with Dresselhaus interaction term
$H_{\alpha}=\pm \frac{1}{2}\alpha(\sigma_1P_1-\sigma_2P_2)$
\subsection{Case-3}
Choosing $\vec A=i\alpha f(r)\vec{r}/r=\vec B^*$ we get
\begin{eqnarray}
 \left(\begin{array}{cc}
   H_0+H_{1} & 0 \\
   0 & H_0+H_{2}
    \end{array} \right)\Psi=E\Psi
\end{eqnarray}
where
\begin{eqnarray}
H_0&=&\frac{1}{2} P^2 \bf {1_{2\times 2}}\nonumber\\
H_1&=&\frac{1}{2}[2\alpha\frac{f(r)}{r}\sigma_jL_j+\alpha^2f^2(r)+\alpha (f^{\prime}(r)+\frac{2f(r)}{r})]\nonumber\\
H_2&=&\frac{1}{2}[-2\alpha\frac{f(r)}{r}\sigma_jL_j+\alpha^2f^2(r)-\alpha (f^{\prime}(r)+\frac{2f(r)}{r})]
\end{eqnarray}
For $f(r)=1/r$
\begin{eqnarray}
H_1=\frac{\alpha}{r^2}\sigma_jL_j+\frac{1}{2}\alpha(\alpha+1)\frac{1}{r^2}
\end{eqnarray}

\begin{eqnarray}
H_2=-\frac{\alpha}{r^2}\sigma_jL_j+\frac{1}{2}\alpha(\alpha-1)\frac{1}{r^2}
\end{eqnarray}

These operators resemble the classical Calogero-Sutherland model with spin-orbit coupling term. $\psi$ and $\eta$ represent the two component spinor states corresponding to attractive and repulsive type coupling respectively.
\par For $f(r)=r$
\begin{eqnarray}
H_1=\frac{1}{2}[2\alpha \sigma_jL_j+3\alpha+\alpha^2 r^2]\nonumber\\
H_2=\frac{1}{2}[-2\alpha \sigma_jL_j-3\alpha+\alpha^2 r^2]
\end{eqnarray}
These operators resemble the Hamiltonian for three dimensional harmonic oscillator with spin-orbit coupling, both attractive and repulsive type.
\subsection{Supersymmetric Harmonic Oscillator}
For supersymmetric harmonic oscillator in one dimension
\begin{eqnarray}
\sigma_1(P_1+i\omega X_1)\psi +2i\eta=0\nonumber\\
\sigma_1(P_1-i\omega X_1)\eta-iE\psi=0
\end{eqnarray}
The corresponding Schr$\ddot{o}$dinger equation can be given by
\begin{eqnarray}
 \left(\begin{array}{cc}
   H_0+\omega/2 & 0 \\
   0 & H_0-\omega/2
    \end{array} \right)\Psi=E\Psi
\end{eqnarray}
Where $H_0=\frac{1}{2}(P^2+\omega^2X^2)$ and the Schrodinger equations corresponding to $\psi$ and $\eta$ represent the fermionic and bosonic sectors respectively of a supersymmetric Hamiltonian.
It is to be noted that the origin of spin orbit coupling is often due to choice of Clifford generalized momenta in the corresponding L$\acute{e}$vy-Leblond type equation.

\section{Conclusion}
The choice of generalized momentum operators most often in Clifford algebraic basis leads to various Hamiltonians of physical interest.The existence of potential terms accompanied by constant or spatially varying L-S coupling terms are basically  a consequence of this choice justifying Hertz contention of kinetic origin of potential even in quantum mechanics with spinor states. The Cliffordization of momentum operator can be expected to produce higher component states for various choices of Clifford bases. A non-hermitian $PT$-symmetric extension of the harmonic oscillator Hamiltonian with real spectrum is hoped to be possible following the same method.
\section{Acknowledgement}
A. C. wishes to thank his colleague Dr. Baisakhi Mal for her valuable assistance in preparing the latex version.

\section{References:}
\par[1]  P. A. M. Dirac: Proc. R. Soc. Lond. A \textbf{117}(1928) 610\\

\par [2] Lectures on Quantum Field Theory:A. Das. World Scientific(2008).\\

\par [3] The Dirac Equation:Bernd Thaller. Springer-Verlag(1992).\\

\par [4] J. M. Levy-Leblond:Commun. Math. Phys.\textbf{6}(1967)286-311.\\

\par [5] Spinors in Physics:J. Hladik.Springer Science+business media, LLC(1999).\\

\par [6] C. R. Hagen:arxiv:1409-5101v1[quant-ph]17 sep 2014.\\

\par [7] Clifford Algebra and Spinors:P. Lounesto. Cham. Univ. Press(2001).\\

\par [8] E. I. Rashba:Sov. Phys. Solid State\textbf{2}(1960)1224-1238.\\

\par [9] Yun-Chang Xiao, Wen-Ji Deng:Superlattices and Microstructures\textbf{48}(2010)181-189.\\

\par [10] Jian-Duo Lu:Physica E\textbf{43}(2010)142-145.\\

\par [11] Xu Yan, Qiang Gu:Solid State Communication\textbf{187}(2014)68-71.\\

\par [12] Quantum Mechanics:Franz Schwabl. Springer(2007).\\

\par [13] The Variational Principle in Mechanics:C.Lanczos.Dover(1986)\\

\par [14] Classical Mechanics(Hamiltonian and Lagrangian Formalism):Alexei Deroglazov. Springer(2010).

\end{document}